\documentclass[aps,prl,twocolumn,superscriptaddress]{revtex4}
\usepackage{graphicx}
\usepackage{mathrsfs}
\usepackage{amsmath}

\begin{document}



\title{Magneto-transport in Disordered Graphene:\\from Weak Localization to Strong Localization}

\author{J. Moser}
\affiliation{CIN2 (CSIC-ICN) Barcelona, Campus UAB, E-08193
Bellaterra, Spain}

\author{H. Tao}
\thanks{Present address: Research Institute of Micro/Nano
Science and Technology, Shanghai Jiao Tong University, 200240,
Shanghai, China.}
\affiliation{CIN2 (CSIC-ICN) Barcelona, Campus UAB,
E-08193 Bellaterra, Spain}

\author{S. Roche}
\affiliation{CIN2 (CSIC-ICN) Barcelona, Campus UAB, E-08193
Bellaterra, Spain}\affiliation{CEA, INAC, SP2M, L\_sim, 17 avenue
des Martyrs, 38054 Grenoble (France)}

\author{F. Alsina}
\affiliation{CIN2 (CSIC-ICN) Barcelona, Campus UAB, E-08193
Bellaterra, Spain}

\author{C. M. Sotomayor Torres}
\affiliation{CIN2 (CSIC-ICN) Barcelona, Campus UAB, E-08193
Bellaterra, Spain}\affiliation{ICREA, 08010 Barcelona, Spain}

\author{A. Bachtold}
\affiliation{CIN2 (CSIC-ICN) Barcelona, Campus UAB, E-08193
Bellaterra, Spain}




\begin{abstract}
We present a magneto-transport study of graphene samples into which
a mild disorder was introduced by exposure to ozone. Unlike the
conductivity of pristine graphene, the conductivity of graphene
samples exposed to ozone becomes very sensitive to temperature: it
decreases by more than 3 orders of magnitude between 100~K and 1~K.
By varying either an external gate voltage or temperature, we
continuously tune the transport properties from the weak to the
strong localization regime. We show that the transition occurs as
the phase coherence length becomes comparable to the localization
length. We also highlight the important role of disorder-enhanced
electron-electron interaction on the resistivity.
\end{abstract}

\maketitle

\section{Introduction}

Quantum interference phenomena in graphene are of fundamental
interest \cite{McCann2009}. A case in point is the localization of
charges, which is a manifestation of two important properties of
this material: First, graphene hosts chiral Dirac fermions. Second,
these fermions reside in two inequivalent valleys at the K and
K$^\prime$ points of the first Brillouin zone. Traveling paths that
are relevant to localization phenomena are phase coherent closed
loops. Because of its chirality, a Dirac fermion residing in a given
valley acquires a phase of $\pi$ upon completion of one loop, which
gives rise to destructive interference with its time-reversed
counterpart. Chirality therefore lowers the probability for
returning paths, and favors weak antilocalization. Restoring
constructive interferences requires inter-valley scattering events
(fermions in the K and K$^\prime$ valleys have opposite
chiralities). This in turn favors weak localization.

\begin{figure}[t]
\includegraphics{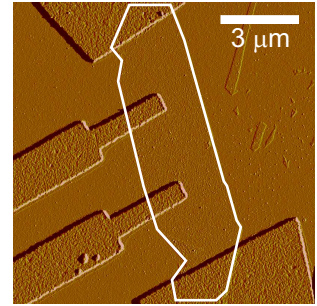}
\caption{(a) Atomic Force Microscopy image of the device contacted
with 4 electrodes. The white contour highlights the sample shape.}
\end{figure}

These quantum interference effects have been actively studied both
theoretically
\cite{Suzuura_Ando_JPSJ,Aleiner,Altland,Mirlin,McCann_PRL06,Morpurgo}
and experimentally
\cite{Geim,deHeer,Savchenko2008,Savchenko2009,Mason}. It has been
predicted that weak localization correction to the semi-classical
(Drude) conductivity will dominate the weak antilocalization
correction as the temperature is lowered, driving graphene to the
strong localization regime \cite{Aleiner}. Surprisingly however,
these corrections measured in graphene samples have remained modest,
even at milliKelvin temperatures.

A clear strategy to study the transition between weak localization
and strong localization is to enhance intervalley scattering. This
can be achieved by introducing short-range scatterers, such as weak
point disorder or lattice defects that result in midgap states
\cite{das_Sarma,Ando,Guinea}. Recently, defect scattering centers
were introduced in graphene using Ne and He ion irradiation, but the
conductivity at the Dirac point remained above $e^2/h$ even down to
cryogenic temperature ($e^2/h$ is the conductivity value for which
the weak localization regime is expected to cross over to the strong
localization one \cite{Fuhrer_PRL09}).

\begin{figure}[t]
\includegraphics{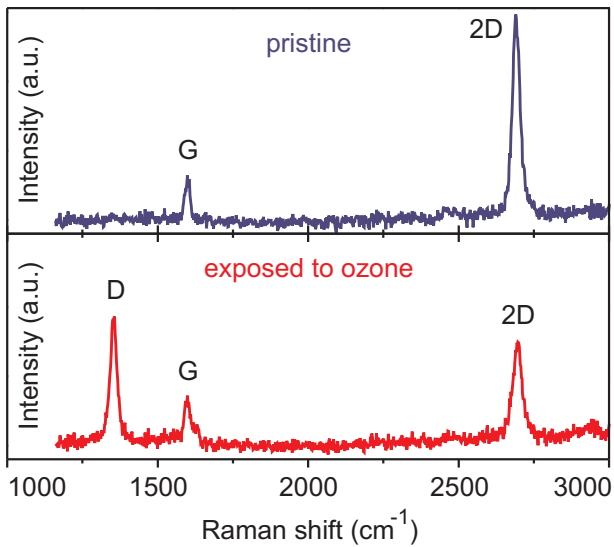}
\caption{Raman spectrum of the device shown in Fig.~1 before (top)
and after (bottom) exposure to ozone. G and 2D modes are shown;
defects created by ozone reveal themselves as a strong D mode. Both
panels have the same intensity axis.}
\end{figure}

Approaches to drive the metallic phase of graphene to an insulator
with an energy band gap have also been explored. In the case of
graphane \cite{Geim_Science09}, $\textrm{sp}^2$ bonds were partially
transformed into $\textrm{sp}^3$ by hydrogenation. There, the
resistivity was found to diverge at low temperatures, in accordance
with the two-dimensional variable range hopping model. The
measurements were interpreted as the result of a modified graphene
that consists of two phases, regions with $\textrm{sp}^3$
hybridization interspersed with $\textrm{sp}^2$ regions
\cite{Geim_Science09}. Similar results were obtained with oxidized
graphene \cite{Gomez-Navarro_NL07,Kaiser_NL09}. Another work
reported transport measurements deep in the strong localization
regime using graphene modified with hydrogen atoms \cite{Bostwick}.

Overall, these transport studies on intentionally disordered
graphene focused either on the metallic regime \cite{Fuhrer_PRL09}
or on the deep localization regime, where the material behaves as an
insulator
\cite{Geim_Science09,Gomez-Navarro_NL07,Kaiser_NL09,Bostwick}. The
possible transition between weak localization and strong
localization remains to be explored.

In this work, we report on a detailed magneto-transport study
unveiling quantum localization effects in disordered graphene.
Disorder is created by exposing graphene to ozone, which introduces
sp$^3$-type defects. By varying either an external gate voltage or
temperature, we continuously tune the transport properties from the
weak localization to the strong localization regime. We show that
the transition occurs as the phase coherence length becomes
comparable to the localization length. In addition, we evidence a
marked contribution of electron-electron interaction to the
resistivity. (This latter effect is a correction to the density of
states and its origin differs from the one of weak localization,
which stems from a modification of the diffusion constant
\cite{Beenakker}).

\section{Experimental results}

Before creating defects in graphene, we first fabricate high-quality
devices using conventional nanofabrication techniques (Fig.~1)
\cite{APL_cleaning}. We mechanically exfoliate graphene from a flake
of Kish graphite on a Si wafer coated with 300~nm of thermal oxide.
We pattern Cr/Au electrodes in a four-point configuration using
electron-beam lithography. We carry out Raman spectroscopy
\cite{ferrari} and low temperature transport measurements to verify
that single layer graphene sheets are of good quality. For the
device discussed in the paper, the D peak is absent before ozone
treatment (Fig.~2, upper panel). In addition, the mobility is
$5500~\textrm{cm}^2/\textrm{Vs}$, and the conductivity at the Dirac
point reaches $5~e^2/h$ and is temperature independent down to
liquid helium temperatures (not shown).

We introduce defect using an ozone treatment, which is a chemically
reactive process known to alter the underlying $\textrm{sp}^2$
network of graphitic systems \cite{Lee_JPC09}. Specifically, we
first clean graphene samples by placing them in a flow of Ar/H$_2$
gas at $300^{\circ}\textrm{C}$ for 3 hours. We then expose the
samples to ozone in a Novascan ozone chamber, where ozone is
produced by ultra-violet irradiation of O$_2$ gas ($\sim 7$~min,
4~atm). The appearance of the D peak in the Raman spectrum (Fig.~2,
lower panel) and the decrease of the mobility down to
$390~\textrm{cm}^2/\textrm{Vs}$ (see below) signal the creation of
additional defects. However, we make sure that the ozone treatment
is mild enough to preserve the crystalline integrity of graphene, as
evidenced by the presence of the G peak and a well defined 2D peak
in Raman spectroscopy. In addition, the elastic mean-free path of
electrons estimated from the Drude conductivity is at least 3~nm
(see below), which is more than one order of magnitude larger than
the C-C bond length.

\begin{figure}[t]
\includegraphics{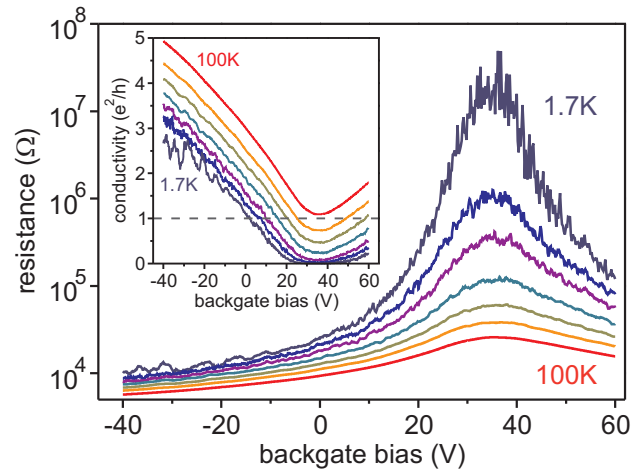}
\caption{(a) 4-point resistance and (b) 4-point conductivity as a
function of backgate bias $V_{g}$. Temperatures are $T=1.7$, 3, 5,
10, 20, 40 and 100~K.}
\end{figure}

This ozone treatment has a large impact on the transport properties
of graphene. The conductivity $\sigma$ becomes very sensitive to
temperature $T$ (Fig.~3). Even though the conductivity at the Dirac
point remains larger than $e^2/h$ at the highest temperature
(100~K), it is reduced by more than 3 orders of magnitude at 1.7~K.
The device behaves as an insulator, at least at low temperature and
in the vicinity of the Dirac point (where the conductivity $\sigma$
as a function of backgate bias $V_{g}$ is lowest, see inset to
Fig.~3).

\begin{figure*}[t]
\includegraphics{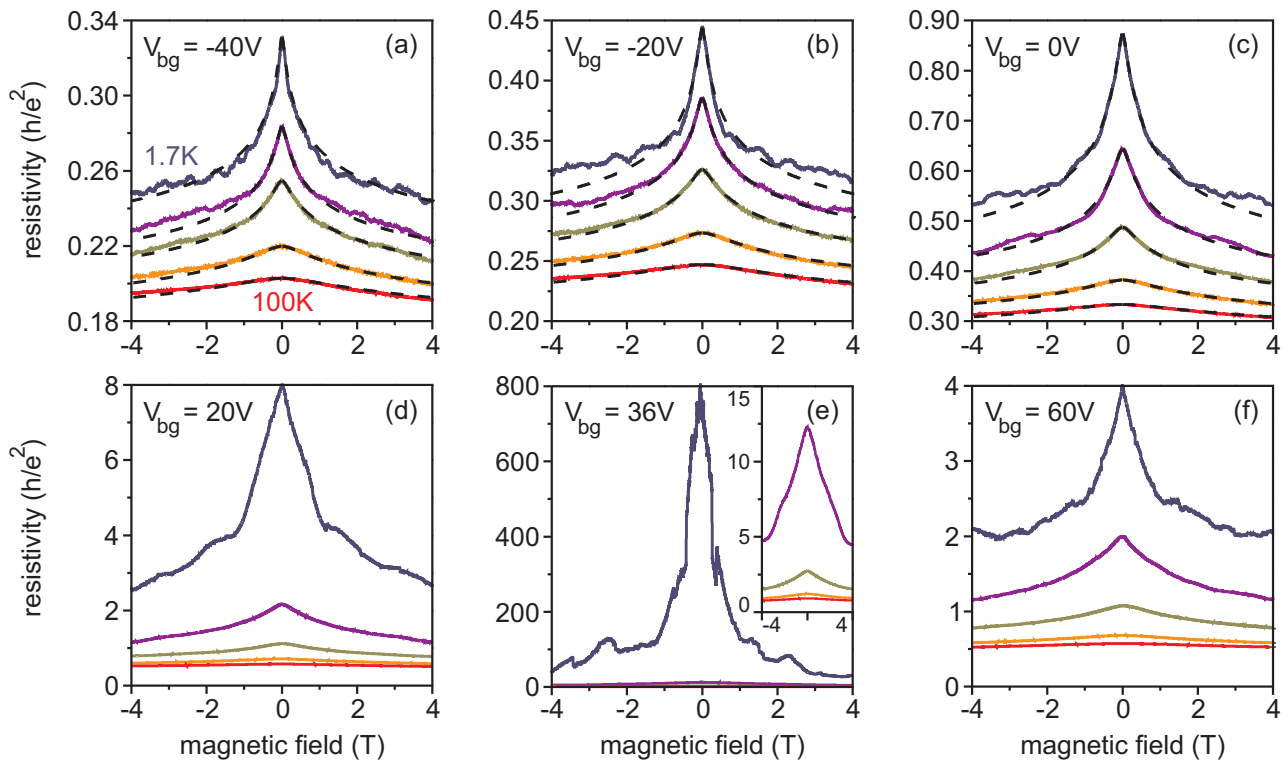}
\caption{Resistivity as a function of magnetic field, for $T=1.7, 5,
15, 50$ and 100~K. Fits to eq.~(1) are shown by dashed lines away
from the Dirac point and for conductivity values larger than
$e^{2}/h$. Inset to Fig.~4e displays the resistivity at the Dirac
point at 5, 15, 50 and 100~K (top to bottom).}
\end{figure*}

To reveal the contributions of quantum interferences, we explore the
transport properties in the presence of a magnetic field $B$
(Fig.~4a-f). In all cases, the magnetoresistance is negative, and
the resistivity changes with $B$ by an amount that strongly depends
on $V_{g}$. This change becomes increasingly large as the density
approaches the Dirac point ($V_{g}=36~\textrm{V}$, Fig.~4e) whereas
it remains moderate at high charge density $n$
($V_{g}=-40~\textrm{V}$, Fig.~4a).

\section{Discussion}

Both the low temperature insulating behavior \cite{Aleiner,Mirlin}
and the sharp D peak in Raman spectra show that the ozone treatment
introduces significant intervalley scattering. A way to
quantitatively compare intervalley and intravalley scattering rates
is to examine the $V_{g}$ dependence of the conductivity. Assuming
only weak point disorder and charged impurity disorder, we obtain
that the intervalley and the intravalley scattering times are
comparable, about 10~fs for $V_{g}=-40~{V}$. Conversely, before
ozone treatment the intervalley scattering time is 300~fs and the
intravalley scattering time is 100~fs at the same $n$. To determine
these scattering times, we consider weak point disorder for the
intervalley scattering time
$\tau_{\textrm{inter}}=\sigma_{\textrm{sr}}\cdot h/(2e^{2}\cdot
v_{F}\sqrt{\pi n})$ and charged-impurity disorder for the
intravalley scattering time $\tau_{\textrm{intra}}=h/(2e\cdot
v_{F}\sqrt{\pi})\cdot\mu\cdot\sqrt{n}$. The conductivity
$\sigma_{\textrm{sr}}$ [due to short range scattering] and the
mobility $\mu$ can be extracted from the $\sigma(V_{g})$ measurement
taken at 100~K (assuming that localization effects are vanishingly
small at higher temperature) using
$\sigma^{-1}=\sigma_{\textrm{sr}}^{-1}+(e\mu n)^{-1}$
\cite{Fuhrer_PRL08}. We get $\sigma_{\textrm{sr}}=5\cdot
10^{-3}~\textrm{S}$ and $\mu=5500~\textrm{cm}^{2}/\textrm{Vs}$ for
the pristine graphene, and  $\sigma_{\textrm{sr}}=4.1\cdot
10^{-4}~\textrm{S}$ and $\mu=390~\textrm{cm}^{2}/\textrm{Vs}$ after
ozone treatment. We note that lattice defects resulting in midgap
states were recently identified as a new source of intervalley
scattering \cite{Fuhrer_PRL09}. The latter scattering results in a
linear $n$ dependence of $\sigma$ so its contribution to the
conductivity cannot be discriminated from the one of
charged-impurity disorder in a $\sigma(V_{g})$ measurement. As such,
it can modify $\tau_{\textrm{inter}}$ and $\tau_{\textrm{intra}}$
and the intervalley scattering time of 10~fsec is an upper bound.

The magnetoresistance measurements at high $n$ can be well described
by the weak localization (WL) theory developed for graphene
\cite{McCann_PRL06}. The correction to the semi-classical (Drude)
conductivity reads
\begin{equation}
\begin{aligned}
\delta\sigma_{\textrm{graphene}}=\frac{e^{2}}{\pi h}\left[F\left(\frac{\tau_{B}^{-1}}{\tau_{\phi}^{-1}}\right)-F\left(\frac{\tau_{B}^{-1}}{\tau_{\phi}^{-1}+2\tau_{\textrm{inter}}^{-1}}\right)\right.\\
\left.-2F\left(\frac{\tau_{B}^{-1}}{\tau_{\phi}^{-1}+\tau_{\textrm{inter}}^{-1}+\tau_{\textrm{intra}}^{-1}}\right)\right]
\end{aligned}
\end{equation}
with $F(z)=\ln z + \psi(0.5+z^{-1})$ where $\psi$ is the digamma
function, $\tau_{B}^{-1}=4eDB/\hbar$, and
$\tau_{\phi}=L_{\phi}^{2}/D$ the phase coherence time. To compare
with the experiment, we let the diffusion constant
$D=0.5v_{F}^{2}\cdot(\tau_{\textrm{inter}}^{-1}+\tau_{\textrm{intra}}^{-1})^{-1}$
and consider only weak point disorder for intervalley scattering and
charged-impurity disorder for intravalley scattering. Accordingly,
the phase coherence time $\tau_{\phi}$ is the only fitting parameter
necessary. As illustrated in Fig.~4(a,b,c) (where
$\sigma(B=0)>e^{2}/h$), we find a good agreement between experiments
and theory. A satisfactory agreement is also obtained by comparing
measurements to WL predictions for conventional two-dimensional
metals, which, moreover, yields the same phase coherence time. As
for the magnetoresistance measurements at lower $n$ (Fig. 4(d,e,f)),
the resistivity can change with $B$ by a large amount. Comparing the
measurements to theory is however difficult at this stage. More
measurements at low temperature will be needed to discern among the
various predicted dependencies \cite{xiB}.

\begin{figure}[t]
\includegraphics{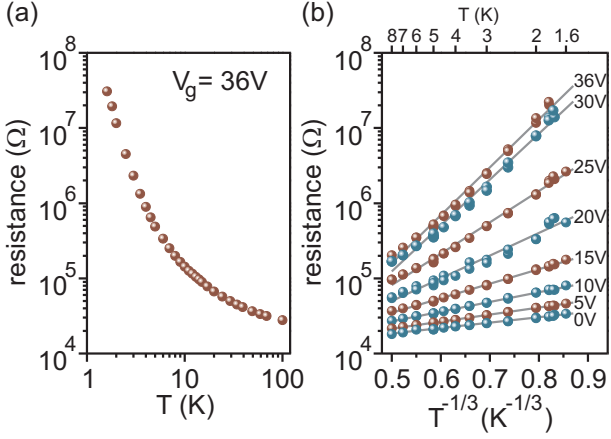}
\caption{(a) Temperature dependence of the resistance at the Dirac
point. (b) At low temperature, the sample resistance is fit to the
two-dimensional variable-range hopping model for several gate biases
in the vicinity of the Dirac point.}
\end{figure}
\begin{figure}[t]
\includegraphics{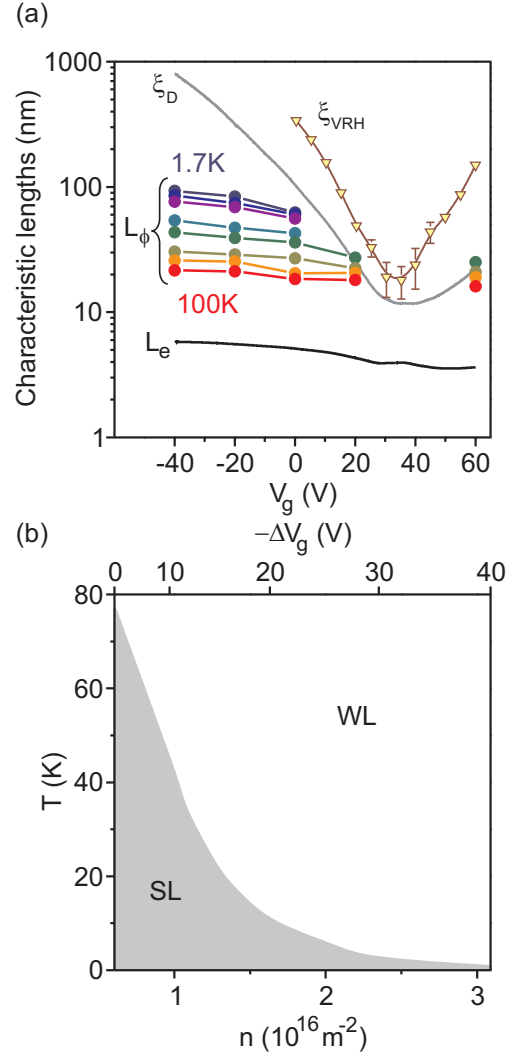}
\caption{(a) Characteristic lengths as a function of backgate bias
$V_{g}$. $L_{e}$ is the elastic mean free path. Coherence lengths
$L_{\phi}$ at $T=1.7$, 3, 5, 15, 25, 50, 70 and 100~K (top to
bottom) are shown. $\xi_{VRH}$ is the localization length estimated
from the variable range hopping model (eq.~2) and $\xi_{D}$ is the
localization length estimated from the scaling theory (eq.~3). (b)
Phase diagram (temperature $T$, charge density $n$) showing the
semi-metallic regime where weak localization (WL) is observed and
the insulator regime where strong localization (SL) prevails. The
top axis shows the corresponding backgate biases $\Delta V_{g}$
measured from the Dirac point.}
\end{figure}

We now turn our attention to the temperature dependence of the
resistance. The resistance at the Dirac point diverges at low
temperature, as expected for an insulating regime (Fig.~5a). Fig.~5b
shows that the temperature-dependent resistance is consistent with
two-dimensional variable-range hopping, $R\propto
\exp((T_{0}/T)^{1/3})$. Even though the measurement could also be
described by a simple thermal activation behavior, for the time
being we will restrict our analysis to a variable range hopping
scenario, which is the conventional mechanism to describe low
temperature conduction in strongly disordered materials
\cite{Geim_Science09,Gomez-Navarro_NL07,Kaiser_NL09,Efros}. This
allows us to extract the localization length $\xi_{VRH}$ from the
fitting parameter $T_{0}$ \cite{Efros}:
\begin{equation}
\xi_{VRH}=\sqrt{\frac{13.8}{k_{B}\rho T_{0}}}
\end{equation}
where $\rho$ is the density of states of graphene. To estimate
$\rho$, we assume that $n=C_{g}\sqrt{V_{g}^{2}+V_{0}^{2}}/e$ to take
the fluctuations of the Dirac point with respect to the Fermi energy
into account \cite{Amelia_PRL}. Here, $C_{g}=1.15\cdot
10^{-4}$~F/m$^{2}$ is the backgate capacitance \cite{Amelia_PRL}.
Letting $V_{0}=8~\textrm{V}$ (which roughly accounts for the
smoothing of the $\sigma$ \textit{vs} $V_{g}$ curve around the Dirac
point at the highest temperature), we obtain
$\xi_{VRH}=15~\textrm{nm}$ at the Dirac point. Away from the Dirac
point, we find that $\xi_{VRH}$ increases upon increasing $n$
(Fig.~6a). As a comparison, we can evaluate the localization length
using the rough estimate \cite{Gershenson_PRL}:
\begin{equation}
\xi_D\simeq
L_{e}\exp\left(\frac{\sigma_{\textrm{D}}}{e^{2}/h}\right)
\end{equation}
The elastic length $L_{e}$ is derived from the Drude conductivity
$\sigma_{\textrm{D}}$ as $L_{e}=v_F\sigma_{\textrm{D}}\cdot
h/(2e^{2}v_{F}\sqrt{\pi n})$, assuming that $\sigma_{\textrm{D}}$ is
the conductivity measured at a temperature of 100~K. This yields
$\xi_D\simeq 12~\textrm{nm}$ at the Dirac point, which is close to
the variable range hopping estimate. The agreement is also quite
reasonable at higher $n$ (Fig.~6a). Overall, this rough agreement
gives us confidence in the order of magnitude of the localization
length.

The transition from weak localization to strong localization can be
understood by contrasting the fundamental transport length scales
extracted from our experimental data (Fig.~6a). As long as the phase
coherence length $L_{\phi}$ remains smaller than the estimated
localization length $\xi$, the weak localization regime prevails.
Whenever $L_{\phi}$ becomes comparable to $\xi$, we observe that the
conductivity is close to $e^{2}/h$, the value at which the
transition to strong localization is expected \cite{Lee_RMP85}.
Fig.~6a displays $L_{\phi}$ only when $\sigma>e^{2}/h$ at zero
magnetic field (and except at the Dirac point). In the opposite case
$(\sigma<e^{2}/h)$, the comparison between experiment and
weak-localization theory becomes worse, so that extracting a value
for $L_{\phi}$ is meaningless.

Graphene offers the possibility to tune the carrier density $n$ with
$V_{g}$, which provides a practical knob to test the localization
theory. From our measurements, we can construct a "phase diagram" of
the transition from weak (WL) to strong localization (SL) as a
function of $n$ and temperature (Fig.~6b). To do this, we use the
temperature dependence of $\sigma$ (a few such curves are shown in
Fig.~3(b)) and define the transition as $\sigma=e^{2}/h$. We
emphasize that the transition is expected to be gradual and to
develop as a smooth crossover \cite{Lee_RMP85}. Figure 6b shows that
the WL-SL transition is very sensitive to $n$ at low carrier
concentration. This is because of the strong variation of $\xi$ upon
varying $n$ (Fig. 6a). By contrast, the other fundamental transport
lengths, $L_e$ and $L_{\phi}$, remain essentially constant.

Eventually, one interesting outcome of our measurements is that they
reveal the importance of Coulomb interaction between charge
carriers. This can be seen in the magnetoresistance measurements at
high magnetic field (Fig.~4a) where the conductivity at
$B=4~\textrm{T}$ is definitely temperature dependent. The
contribution of weak localization is reduced at high magnetic field
and the remaining correction to $\sigma$ is usually attributed to
Coulomb interaction. The correction to the conductivity due to
Coulomb interaction in a two-dimensional metal reads
\cite{Beenakker,AAL}
\begin{equation}
\delta\sigma_{ee}=-\frac{e^{2}}{2\pi^{2}\hbar}\cdot g_{2D}\cdot
\ln\left(\frac{\hbar}{k_{B}T\tau_{e}}\right)
\end{equation}
where $g_{2D}$ is a constant of the order of unity. Figure 7 shows
that experiment agrees well with the theory using $g_{2D}=0.8$.
Unlike weak localization, which originates from a change in the
diffusion constant \cite{ Beenakker}, this Coulomb interaction
effect is a correction to the density of states. These results
indicate that the correction to the conductivity due to Coulomb
interaction cannot be neglected in the transition between weak
localization and strong localization in graphene, as it is treated
in existing theoretical works
\cite{Suzuura_Ando_JPSJ,Altland,Aleiner,Nomura,Mirlin,Robinson,Roche_PRL,Morpurgo,McCann_PRL06}.

\begin{figure}[t]
\includegraphics{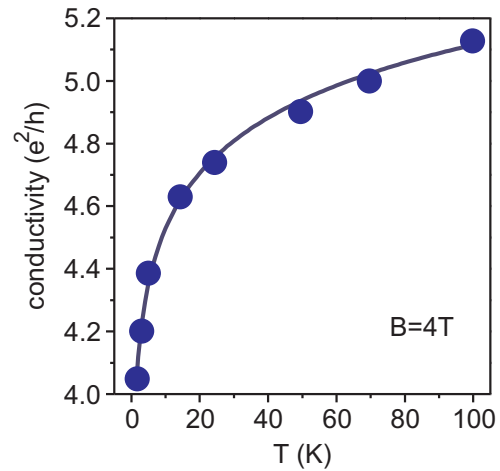}
\caption{Temperature dependence of the conductivity at $B=4$~T
(dots) and fit to equation (4) (solid line).}
\end{figure}

\section{Conclusion}

We have reported on the crossover from weak localization to strong
localization in disordered graphene, the disorder being created with
ozone. For this, we have carried out magneto-transport measurements
as a function of gate voltage and temperature. We have shown that
the transition between weak and strong localization occurs as the
phase coherence length becomes comparable to the localization length
and that the transition is very sensitive to the charge carrier
concentration. In addition, we have demonstrated the importance of
the resistivity correction due to disorder-induced electron-electron
interaction.

Previous works showed that disorder in graphene obtained by
hydrogenation or oxidation can open an energy gap
\cite{Geim_Science09,Gomez-Navarro_NL07,Kaiser_NL09}. By contrast,
recent calculations reported that this is not the case for graphene
exposed to ozone \cite{nicolas}. In addition, even if such a gap was
created, our measurements suggest that it would be very small.
Indeed, we obtain an energy gap of 1~meV when we compare the
temperature dependence of the resistivity at the Dirac point to a
thermal activation behavior. This low value suggests that a gap, if
it exists, would be relevant only to measurements near the Dirac
point (in our case, it would be included in the interpretation of
the measurements at $V_{g}=36$~V). Further experimental work is
needed to demonstrate whether this energy gap exists or not, using
e.g. scanning tunneling microscopy techniques.

\section{Acknowledgements}

We are grateful to M. Lira-Cantu for the use of the ozone chamber
and acknowledge fruitful discussion with F. Guinea. This work was
supported by a EURYI grant, the EU grant FP6-IST-021285-2, and the
MICINN grants FIS2008-06830 and FIS2009-10150. S. R. acknowledges
the ANR/P3N2009 (NANOSIM-GRAPHENE project number
ANR-09-NANO-016-01).\\

\end{document}